\documentclass[paper]{JHEP3}
\usepackage{epsfig}
\usepackage{bbm}

\def\sla#1{\ifmmode%
\setbox0=\hbox{$#1$}%
\setbox1=\hbox to\wd0{\hss$/$\hss}\else%
\setbox0=\hbox{#1}%
\setbox1=\hbox to\wd0{\hss/\hss}\fi%
#1\hskip-\wd0\box1 }

\title{Gluon-induced $WW$ background to Higgs boson searches
 at the LHC}

\author{T.~Binoth\\
Institut f\"ur Theoretische Physik, Universit\"{a}t W\"{u}rzburg, 
D-97074 W\"{u}rzburg, Germany}

\author{M.~Ciccolini\\
Paul Scherrer Institut, CH-5232 Villigen PSI, Switzerland}

\author{N.~Kauer\\
Institut f\"{u}r Theoretische Physik E, RWTH Aachen, D-52056
Aachen, Germany}

\author{M.~Kr\"{a}mer\\
Institut f\"{u}r Theoretische Physik E, RWTH Aachen, D-52056
Aachen, Germany}

\abstract{ Vector-boson pair production is an important background for
  Higgs boson and new physics searches at the Large Hadron Collider
  LHC. We have calculated the loop-induced gluon-fusion process $gg
  \to WW \to {\rm leptons}$, allowing for arbitrary invariant masses
  of the intermediate $W$ bosons. This process contributes at ${\cal
    O}(\alpha_{\rm s}^2)$ relative to quark-antiquark annihilation,
  but its importance is enhanced by the large gluon flux at the LHC
  and by experimental cuts employed in Higgs boson searches.  We find
  that $gg \to WW$ provides only a moderate correction (ca.\ 5\%) to
  the inclusive $W$-pair production cross section at the LHC.
  However, after taking into account realistic experimental cuts, the
  gluon-fusion process becomes significant and increases the
  theoretical $WW$ background estimate for Higgs searches in the $pp
  \to H \to WW \to {\rm leptons}$ channel by approximately 30\%.  }


\preprint{PITHA 05/02\\
          PSI-PR-05-03\\
          WUE-ITP-2005-002}

\begin{document}


\section{Introduction\label{intro-section}}
The search for Higgs particles and for physics beyond the Standard
Model (SM) is central to the physics program at the future Large
Hadron Collider (LHC) at CERN. One of the most promising processes for
Higgs-boson discovery in the mass range $M_{\rm Higgs}$ between 140
and 180~GeV is the $pp \to H \to W^{\ast}W^{\ast} \to
\ell\bar{\nu}\bar{\ell'}\nu'$
channel~\cite{Dittmar:1996ss,Dittmar:1996sp}, where $pp\to H$ proceeds
through gluon-gluon fusion.\footnote{Another favourable Higgs
  discovery channel is the $H\to WW$ decay mode in weak boson fusion,
  $pp \to H(\to WW)+qq$~\cite{Kauer:2000hi}.}  The dominant
irreducible background to this process is $W$-pair continuum
production, which has a cross section times branching ratio more than
five times larger than the Higgs-boson signal. An accurate theoretical
prediction for the background process is crucial to fully exploit the
$pp \to H \to W^{\ast}W^{\ast} \to \ell\bar{\nu}\bar{\ell'}\nu'$
discovery channel, in particular as no Higgs mass peak can be
reconstructed from leptonic $W$ decays.  Besides its significance as a
background to Higgs searches, vector-boson pair production is an
important testing ground for the non-Abelian structure of the SM in
general. It furthermore provides a background for many new physics
searches.

In this paper we present a calculation of the gluon-induced
contribution $gg \to W^{\ast}W^{\ast} \to
\ell\bar{\nu}\bar{\ell'}\nu'$ which proceeds through a quark loop as
shown in Fig.~\ref{fig:feyn}. Although suppressed by two powers of
$\alpha_{\rm s}$ relative to quark-antiquark annihilation,
gluon-fusion is a potentially significant source of $W$-boson pairs at
the LHC because of the large gluon-gluon luminosity.  Furthermore, the
importance of gluon-gluon induced background processes is enhanced by
the experimental Higgs search cuts which exploit the longitudinal
boost and the spin correlations of the $WW$ system to suppress
$W$-pair continuum production through quark-antiquark annihilation.

The hadronic production of $W$ pairs has been studied extensively in
the literature. A recent review can be found in
Ref.~\cite{Haywood:1999qg}. Next-to-leading order (NLO) QCD
corrections to $q\bar{q} \to WW$ have been presented in
Refs.~\cite{Ohnemus:1991kk,Frixione:1993yp}, while NLO calculations
for $q\bar{q} \to WW \to \ell\bar{\nu}\bar{\ell'}\nu'$ including spin
and decay angle correlations can be found in
Refs.~\cite{Ohnemus:1994ff,Dixon:1998py,Dixon:1999di,Campbell:1999ah}.
At large $WW$ invariant masses, also electroweak corrections become
important~\cite{Accomando:2004de}.  The gluon-gluon induced
contribution to on-shell $W$-pair production, $gg \to WW$, has first
been calculated in Ref.~\cite{Glover:1988fe} for the case of massless
quarks circulating in the loop. The calculation has been extended in
Ref.~\cite{Kao:1990tt} to include the top-bottom massive quark loop.
In this paper we present the first calculation of the gluon-induced
process $gg \to W^{\ast}W^{\ast} \to \ell\bar{\nu}\bar{\ell'}\nu'$,
including spin and decay angle correlations and allowing for arbitrary
invariant masses of the intermediate $W$ bosons. The main purpose of
the calculation is to study the impact of gluon-gluon induced
backgrounds to Higgs boson searches in the $H \to W^{\ast}W^{\ast} \to
\ell\bar{\nu}\bar{\ell'}\nu'$ decay mode.  In a first step, and in
order to quantify the general importance of the gluon-fusion process,
we restrict our calculation to the leading contribution which arises
from intermediate light quarks of the first two generations.
Moreover, we do not take into account gluon-gluon induced tree-level
processes of the type $gg \to WW q\bar q$, which have been found to be
strongly suppressed in hadronic $WZ$, $W\gamma$ and $Z\gamma$
production~\cite{Adamson:2002jb}. We finally note that a calculation
similar to ours has been performed for the process $gg \to
Z^{\ast}Z^{\ast} \to 4 l^{\pm}$ in Ref.~\cite{Zecher:1994kb}.

Our calculation demonstrates that the gluon-fusion contribution to
on-shell $W$-pair production only provides a $5\%$ correction to the
inclusive $W$-pair production cross section at the LHC. This result is
consistent with analyses carried out for the processes $gg \to WZ$,
$W\gamma$ and $Z\gamma$~\cite{Adamson:2002jb,Adamson:2002rm}.
However, after taking into account realistic experimental cuts, the
process $gg \to W^{\ast}W^{\ast} \to \ell\bar{\nu}\bar{\ell'}\nu'$
becomes sizeable and enhances the theoretical $WW$ background estimate
for Higgs searches by about $30\%$.  In the following we will present
some details of the calculation and a discussion of the numerical
results.


\section{Calculation \label{calc-section}}
Our aim is to compute gluon-gluon induced off-shell $W$-boson pair
production with decay into lepton-neutrino pairs:
\begin{eqnarray}
\label{amp}
g(p_1,\lambda_1) + g(p_2,\lambda_2) &\to& W^{+*}(p_3) 
 + W^{-*}(p_4)  \nonumber\\
&\to& \bar{\ell'}(p_5,+) + \nu'(p_6,-) + \bar{\nu}(p_7,+) +
\ell(p_8,-) \, ,
\end{eqnarray}
where the momenta and helicities are given in brackets, and where
$p_3=p_5+p_6$ and $p_4=p_7+p_8$.  Generic Feynman diagram topologies
are shown in Fig.~\ref{fig:feyn}.  As mentioned in the introduction,
we allow for arbitrary invariant masses of the $W$ bosons, but focus
on the dominant contribution coming from massless intermediate quark
loops.  This simplifies the computation in two respects.  First, the
restricted number of scales allows for a compact representation of the
amplitude, which has proven necessary to achieve a numerically stable
evaluation.  Second, in the limit of massless quark loops all triangle
topologies in Fig.~\ref{fig:feyn} with an intermediate photon or $Z$
boson vanish, and the box topologies form a gauge invariant set. Note
that in the case of massive quarks in the loop the non-resonant terms
are crucial for maintaining gauge invariance.  

The amplitude $\mathcal{M}$ for the off-shell production of two
charged vector bosons has the following structure:
\begin{eqnarray}
\mathcal{M} =
\varepsilon_{1,\mu_1}\varepsilon_{2,\mu_2}
\mathcal{M}^{\mu_1 \mu_2\mu_3\mu_4} 
P_{\mu_3\nu_3}(p_3,M_W) P_{\mu_4\nu_4}(p_4,M_W) J_{3}^{\nu_3}
J_{4}^{\nu_4}\, ,
\end{eqnarray}
with the lepton currents and $W$ propagators given by
\begin{eqnarray}
J_3^{\mu_3} &=& \bar u(p_6) \gamma^{\mu_3} {\textstyle\frac12} 
(1-\gamma_5) v(p_5) \, , \nonumber\\
J_4^{\mu_4} &=& \bar u(p_8) \gamma^{\mu_4} {\textstyle\frac12} 
(1-\gamma_5) v(p_7) \, , \nonumber\\
P^{\mu\nu}(p,M_W) &=& \frac{g^{\mu\nu}}{p^2 - M_W^2 + i\,M_W
  \Gamma_W } \, .
\end{eqnarray}
We have evaluated the Feynman diagrams using two independent methods.
The expressions for the six box graphs have been compared with output
from FeynArts~3.2~\cite{Hahn:2000kx}.  The algebraical manipulations
were performed with {\sc Form}~\cite{Vermaseren:2000nd} and {\sc
  Maple}.

Numerical tensor reduction which is commonly used in standard
algebraic packages is in general not sufficient to achieve a
numerically stable amplitude representation. In these approaches the
problem of inverse Gram determinants is usually avoided by applying
experimental cuts from the start such that kinematically dangerous
regions do not contribute to the cross section.  In the case of
$W$-boson pair production with subsequent leptonic decays, however,
this is not possible, as the invisible neutrinos do not allow to apply
an experimental cut on the transverse momentum of the $W$ bosons to
avoid large denominators. To overcome this problem we have performed a
fully algebraic tensor and scalar integral reduction following the
methods described in Ref.~\cite{Binoth:1999sp}.  The result has been
cross checked using standard methods
\cite{'tHooft:1978xw,Passarino:1978jh}.  Each helicity amplitude has
been grouped into explicitly gauge invariant tensor structures and
expressed in a basis of scalar integrals similar to those defined in
Ref.~\cite{Binoth:2003xk}.  The respective coefficients have been
simplified algebraically using {\sc Maple} in a fully automated way so
that at most one inverse power of a Gram determinant remained.
Finally, the simplified helicity amplitudes were coded in two
independent computer programs. Our representation of the cross section
is numerically stable without any cuts. More details of the
calculation will be presented in a forthcoming article.


\section{Results \label{results-section}}

In this section we present numerical results for the process $pp \to
W^{\ast}W^{\ast}\to \ell\bar{\nu}\bar{\ell'}\nu'$ at the LHC.
We tabulate the total cross section and the cross
section for two sets of experimental cuts.  In addition, we show various
differential distributions. The experimental cuts include a set of
``standard cuts''~\cite{Dixon:1999di}, motivated by the finite
acceptance and resolution of the detectors, where we require all
charged leptons to be produced at $p_{T,\ell} > 20$ GeV and
$|\eta_\ell| < 2.5$, and a missing transverse momentum $\sla{p}_T >
25$~GeV. Cross sections calculated with this set of cuts will be 
labelled $\sigma_{std}$. 

As emphasized in the introduction, $W$-boson pair production
constitutes the dominant irreducible background to Higgs searches in
the $H \to W^{\ast}W^{\ast} \to \ell\bar{\nu}\bar{\ell'}\nu'$ decay
mode, with a cross section times branching ratio more than five times
larger than the Higgs-boson signal.\footnote{Other, less important
  backgrounds include $W$-boson production from top-quark decays.}
Various cuts have been proposed for the experimental searches to
enhance the signal-to-background
ratio~\cite{Dittmar:1996ss,Dittmar:1996sp,DittmarDreinerIII,
  JakobsTrefzger,Green:2000um, Davatz:2004zg}.  We have adopted a set
of cuts similar to those advocated in a recent experimental
study~\cite{Davatz:2004zg}.  In addition to the ``standard cuts''
defined above, we require that the opening angle between the two
charged leptons in the plane transverse to the beam direction should
satisfy $\Delta\phi_{T,\ell\ell} < 45^{\circ}$ and that the dilepton
invariant mass $M_{\ell\ell}$ be less than $35$~GeV.  Furthermore, the
larger and smaller of the lepton transverse momenta are restricted as
follows: $25~\mbox{GeV} < p_{T,{\rm min}}$ and $35~\mbox{GeV} < p_{T,
  {\rm max}} < 50~\mbox{GeV}$.  Finally, a jet-veto is imposed that
removes events with jets where $p_{T,{\rm jet}} > 20$~GeV and
$|\eta_{\rm jet}| < 3$. Cross sections evaluated with the Higgs
selection cuts will be labelled $\sigma_{bkg}$.

To obtain numerical results we use the following set of input
parameters:
\begin{displaymath}
\begin{array}{llllll}
M_W & = 80.419~{\rm GeV}, \quad & 
M_Z & = 91.188~{\rm GeV}, \quad & 
G_\mu & = 1.16639 \times 10^{-5}~{\rm GeV}^{-2}, \\ 
\Gamma_W & = 2.06~{\rm GeV}, & 
\Gamma_Z & = 2.49~{\rm GeV}, &
V_{\rm CKM} & = \mathbbm{1}\, .
\end{array}
\end{displaymath}
The weak mixing angle is given by $c_{\rm w} = M_W/M_Z,\ s_{\rm w}^2
= 1 - c_{\rm w}^2$.  The electromagnetic coupling is defined in the
$G_\mu$ scheme as $\alpha_{G_\mu} = \sqrt{2}G_\mu
M_W^2s_{\rm w}^2/\pi$.  The masses of external fermions are
neglected.  We restrict our calculation to light intermediate fermions
which are treated as massless. The $pp$ cross sections are calculated
at $\sqrt{s} = 14$~TeV employing the CTEQ6L1 and CTEQ6M
\cite{Pumplin:2002vw} parton distribution functions at tree- and
loop-level, corresponding to $\Lambda^{\rm LO}_5 = 165$ MeV and
$\Lambda^{\overline{{\rm MS}}}_5 = 226$ MeV with one- and two-loop
running for $\alpha_s(\mu)$, respectively.  The renormalization and
factorization scales are set to $M_W$.  Fixed-width Breit-Wigner
propagators are used for unstable gauge bosons.

We compare results for $WW$ production in gluon scattering with LO and
NLO results for the quark scattering processes. Since we are
interested in $WW$ production as a background, the $gg\to H \to WW$
signal amplitude is not included.\footnote{In the case of a light,
  narrow-width Higgs boson, interference effects between signal and
  background are expected to be strongly suppressed, see
  Refs.~\cite{Dicus:1987fk,Dixon:2003yb}.}  The LO and NLO quark
scattering processes are computed with MCFM \cite{Campbell:1999ah},
which implements helicity amplitudes with full spin correlations
\cite{Dixon:1998py} and includes finite-width effects and
single-resonant corrections.  Table~\ref{tbl:xsections} shows gluon
and quark scattering cross sections for the LHC.  Total cross sections
($\sigma_{tot}$) are compared with cross sections when standard LHC
cuts ($\sigma_{std}$) and selection cuts optimized for Higgs boson
searches ($\sigma_{bkg}$) are applied (see above for the definition of
the cuts).  The $gg$ process only yields a 5\% correction to the total
$WW$ cross section calculated from quark scattering at NLO QCD.  When
realistic Higgs search selection cuts are applied the correction
increases to 30\%.  We expect that the inclusion of the $t$-$b$ quark
mediated $gg\to WW$ amplitude contribution will lead to a further
increase.  Note that the experimental Higgs search cuts include a jet
veto which suppresses large contributions from gluon-quark scattering
at NLO and thereby reduces the K-factor for $q\bar{q}\to WW$ from 1.6
to 1.1.

Table~\ref{tbl:xsections} also includes an estimate of the remaining
theoretical uncertainty due to the QCD scale variation. We have varied
the renormalization and factorization scales independently between
$M_W/2 \le \mu_{\rm ren,fac} \le 2M_W$. We find that the largest cross
section prediction corresponds to choosing $\mu_{\rm ren} = M_W/2$ and
$\mu_{\rm fac} = 2M_W$, while the reverse combination $\mu_{\rm ren} =
2M_W$ and $\mu_{\rm fac} = M_W/2$ yields the smallest value. For the
gluon-gluon induced contribution this is obvious, as the
renormalization and factorization scale dependence is entirely due to
$\alpha_{\rm s}$ and the parton distribution functions, respectively:
decreasing $\mu_{\rm ren}$ increases $\alpha_{\rm s}$, while
increasing $\mu_{\rm fac}$ increases the gluon-gluon luminosity. The
latter effect turns out to be small compared to the former. For the
$gg \to WW$ process we find a scale uncertainty of approximately
$25\%$. The scale dependence of the NLO quark-scattering contribution
is more intricate and depends sensitively on the cuts employed. For
the $q\bar{q} \to WW$ process we find approximately $5\%$ uncertainty.
The Higgs selection cuts, which suppress the gluon-quark scattering
contribution, further reduce the scale dependence to a level of only
about 2\%. This is a remarkably small change that, we suspect,
underestimates the uncertainty. A more detailed discussion of the
scale uncertainty for $q\bar{q}\to WW$ can be found in
Ref.~\cite{Dixon:1999di}.

The cross section for $pp \to WW \to \ell\bar{\nu}\bar{\ell'}\nu'$ is
proportional to the fourth power of the electromagnetic coupling
$\alpha$. As mentioned above, we have evaluated $\alpha$ in the
$G_\mu$ scheme. The $G_\mu$ scheme is a preferred choice as it absorbs
large universal higher-order electroweak effects in the LO prediction.
Alternatively, $\alpha$ can be identified with the fine-structure
constant $\alpha(0)$ or the running electromagnetic coupling
$\alpha(Q^2)$ at a high energy scale $Q$. Setting $\alpha = \alpha(0)$
[$\alpha = \alpha(M_Z)]$ results in a decrease [increase] of the
cross-section prediction by approximately 10\%.

Off-shell effects reduce the total $gg\to WW$ cross section in narrow
width approximation by 2.6\%. The Higgs search selection cuts amplify
the effect to about $-$6\%.  The quark scattering process is affected
in a similar way.  We conclude that the narrow width approximation
yields conservative estimates for the $pp\to WW$ background. Note that
the Higgs search cuts have been optimized for a Higgs boson mass of
$M_{\rm Higgs} = 165$~GeV. Off-shell effects may have a larger impact
when other cuts, more suitable for lighter Higgs masses $M_{\rm Higgs}
< 2 M_W$, are applied.

We have compared our results with the $gg \to WW$ cross section
calculation presented in Ref.~\cite{Glover:1988fe}. Adopting the
set-up of Ref.~\cite{Glover:1988fe} we find perfect agreement for the
invariant $WW$ mass distribution. Note that the large impact of the
gluon-gluon induced contributions to $W$-pair production at the LHC
suggested in previous analyses \cite{Glover:1988fe,Kao:1990tt} is not
confirmed in our calculation.  The difference is due to using modern
parton distribution functions, which show a less pronounced rise of
the gluon-luminosity towards small values of $x$, and the associated
smaller values of the strong coupling $\alpha_{\rm s}$.

Selected differential distributions for $pp \to W^{\ast}W^{\ast}\to
\ell\bar{\nu}\bar{\ell'}\nu'$ at the LHC are shown in
Figs.~\ref{fig:mll}, \ref{fig:etal} and \ref{fig:delphill}. The
standard set of cuts defined above has been applied throughout. 

Figure~\ref{fig:mll} shows the distribution in the invariant mass of
the pair of charged leptons. We compare the gluon-gluon induced
contribution with the quark scattering process in LO and NLO. We
observe that the invariant mass distribution of the gluon-gluon
induced process is similar in shape to the quark scattering
contributions and suppressed by more than one order of magnitude in
normalization.

$W$-boson pairs produced in quark-antiquark scattering at the LHC are
in general strongly boosted along the beam axis. Gluon induced
processes on the other hand result in $WW$ events at more central
rapidities. This feature is born out by the distribution in the
pseudorapidity of the negatively charged lepton shown in
Fig.~\ref{fig:etal}. In order to distinguish the shapes of the various
contributions we have chosen a linear vertical scale and plot the
gluon-gluon contribution multiplied by a factor~10. Compared to LO
quark-antiquark scattering, the lepton distribution of the gluon-gluon
process shows a more pronounced peak at central rapidities. We also
observe an enhancement of the NLO corrections at central rapidities
which is due to the substantial contribution of gluon-quark processes
at NLO.

Figure~\ref{fig:delphill} finally shows the distribution in the
transverse-plane opening angle of the charged leptons. This observable
reflects the spin correlations between the $WW$ pair and allows one to
discriminate $W$ bosons originating from scalar Higgs decays and $WW$
continuum production~\cite{Nelson:1986ki}. Note that the importance of
the gluon-gluon process is enhanced by the Higgs search selection cuts
which require a small opening angle $\Delta\phi_{T,\ell\ell} <
45^{\circ}$.


\section{Conclusions \label{concl-section}}
We have calculated the loop-induced gluon-fusion process $gg \to
W^{\ast}W^{\ast} \to \ell\bar{\nu}\bar{\ell'}\nu'$ which provides an
important background for Higgs boson searches in the $H \to WW$
channel at the LHC. We have presented numerical results for the total
cross section, the cross section with two sets of experimental cuts
and various differential distributions.  Our calculation demonstrates
that the gluon-fusion contribution to on-shell $W$-pair production
only yields a $5\%$ correction to the inclusive $W$-pair production
cross section at the LHC.  However, after imposing realistic Higgs
search selection cuts, the process $gg \to W^{\ast}W^{\ast} \to
\ell\bar{\nu}\bar{\ell'}\nu'$ becomes the dominant higher-order
correction to the $WW$ background estimate and enhances the
theoretical prediction from quark-antiquark scattering at NLO by
approximately $30\%$. We conclude that gluon-gluon induced $W$-pair
production is essential for a reliable description of the background
and has to be taken into account to exploit the discovery potential of
Higgs boson searches in the $pp\to H\to WW \to {\rm leptons}$ channel
at the LHC.


\acknowledgments A major part of this work was done when all authors
were at the University of Edinburgh. We would like to thank the
Edinburgh Particle Physics Theory Group and the British Research
Council PPARC for support. We are grateful to Herbi Dreiner for
drawing our attention to the importance of gluon-induced backgrounds
to Higgs boson searches. We benefited from discussions with Michael
Dittmar, Herbi Dreiner, and Adrian Signer. Special thanks go to Peter
Marquard and Jochum van der Bij for a comparison of results and to
John Campbell for his help in using the MCFM program.  The work of
T.B. is supported by the Bundesministerium f{\"u}r Bildung und
Forschung (BMBF, Bonn, Germany) under the contract number 05HT4WWA2.
The work of N.K.~is supported in part by the DFG
Sonderforschungsbereich/Transregio 9 ``Computer-gest\"{u}tzte
Theoretische Teilchenphysik''.



\EPSFIGURE[p] {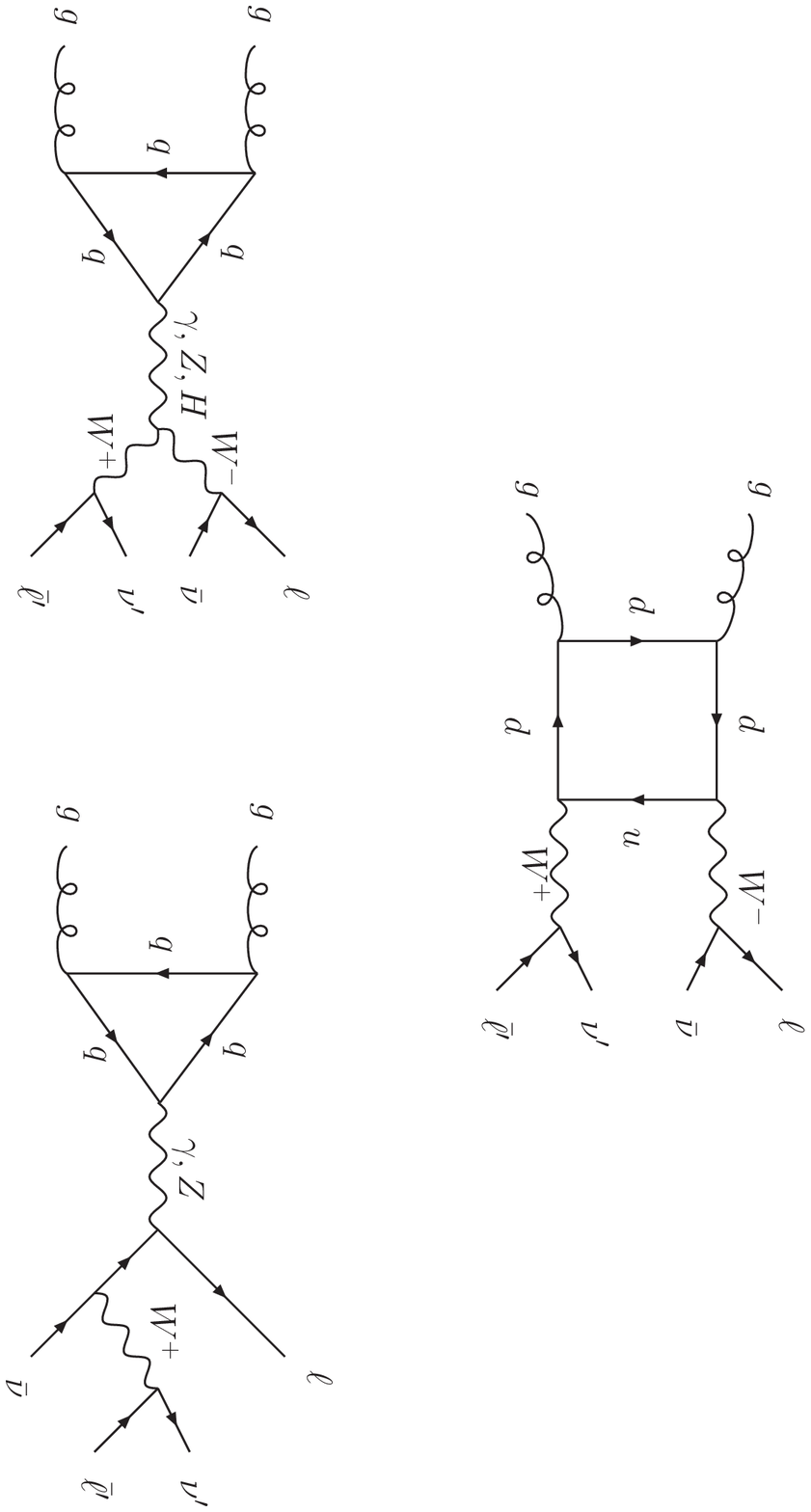,width=7.0cm, angle=90}
{\label{fig:feyn} Generic Feynman diagrams for the process $gg \to
  W^{\ast}W^{\ast} \to \ell\bar{\nu}\bar{\ell'}\nu'$.}


\TABLE[p]{
\renewcommand{\arraystretch}{1.5}
\begin{tabular}{|c|c|cc|c|c|}
 \cline{2-6}
\multicolumn{1}{c|}{} & \multicolumn{5}{c|}{$\sigma(pp \to W^{\ast}W^{\ast}\to
   \ell\bar{\nu}\bar{\ell'}\nu')$~[fb]} \\ \cline{2-6}
\multicolumn{1}{c|}{} & & 
\multicolumn{2}{c|}{\raisebox{1ex}[-1ex]{$q\bar{q}$}}
& \multicolumn{1}{c|}{} &\multicolumn{1}{c|}{} \\[-1.5ex]
\cline{3-4}
\multicolumn{1}{c|}{} & 
\multicolumn{1}{c|}{\raisebox{2.7ex}[-2ex]{$gg$}} & 
\raisebox{0.9ex}{LO} & \raisebox{0.9ex}{NLO} 
& \raisebox{2.7ex}[-2ex]{$\frac{\sigma_{\rm NLO}}{\sigma_{\rm LO}}$} & 
  \raisebox{2.7ex}[-2ex]{$\frac{
 \sigma_{{\rm NLO}+gg}}{\sigma_{\rm NLO}}$}
\\[-1.5ex]
\hline
 $\sigma_{tot}$ & $53.61(2)^{+14.0}_{-10.8}$   & $875.8(1)^{+54.9}_{-67.5}$ &
 $1373(1)^{+71}_{-79}$ & 1.57 & 1.04 \\
 \hline
 $\sigma_{std}$ & $25.89(1)^{+6.85}_{-5.29}$    & $270.5(1)^{+20.0}_{-23.8}$ &
 $491.8(1)^{+27.5}_{-32.7}$ & 1.82 & 1.05 \\
 \hline
 $\sigma_{bkg}$ & $1.385(1)^{+0.40}_{-0.31}$ & $4.583(2)^{+0.42}_{-0.48}$ &
 $4.79(3)^{+0.01}_{-0.13}$ & 1.05 & 1.29 \\
 \hline
\end{tabular}
\vspace*{.5cm}
\caption{\label{tbl:xsections} 
  Cross sections for the gluon and quark scattering contributions to
  $pp \to W^{\ast}W^{\ast}\to \ell\bar{\nu}\bar{\ell'}\nu'$ at the LHC
  ($\sqrt{s} = 14$ TeV) without selection cuts ($tot$), with standard
  LHC cuts ($std$: $p_{T,\ell} > 20$ GeV, $|\eta_\ell| < 2.5$,
  $\sla{p}_T > 25$ GeV) and Higgs search selection cuts ($bkg$, see
  main text) applied. The integration error is given in brackets. The
  errors denote the QCD scale uncertainty obtained by varying
  renormalization and factorization scales independently between
  $M_W/2 \le \mu_{\rm ren,fac} \le 2M_W$. We also show the ratio of
  the NLO to LO cross sections and the ratio of the combined NLO+$gg$
  contribution to the NLO cross section at the central scale $\mu =
  M_W$.}  }


\EPSFIGURE[p] {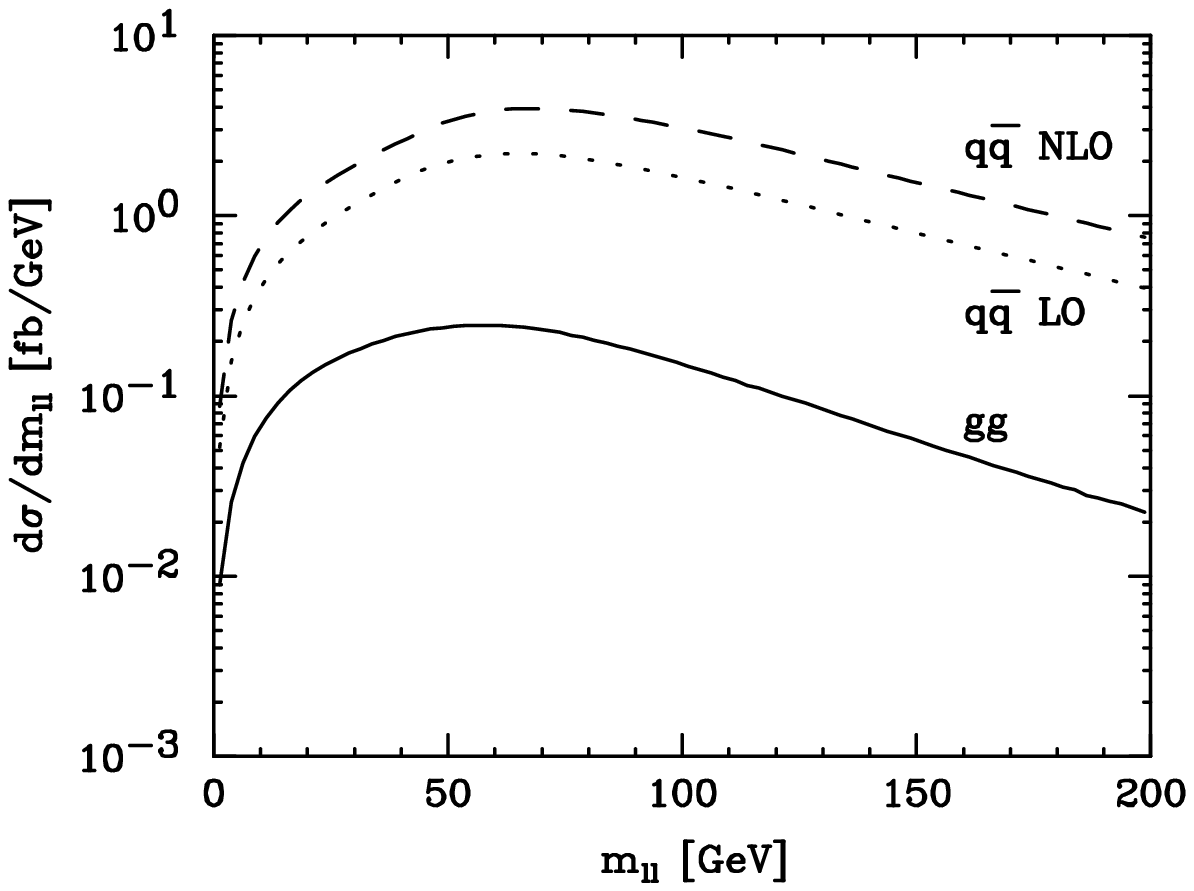,width=12.0cm} {\label{fig:mll}
  Distributions in the charged lepton-pair invariant mass
  $m_{\ell\ell}$ for the gluon scattering process (solid) and the
  quark scattering process in LO (dotted) and NLO QCD (dashed) of
  $pp \to W^{\ast}W^{\ast}\to \ell\bar{\nu}\bar{\ell'}\nu'$ at the
  LHC. Input parameters as defined in the main text. Standard LHC cuts
  have been applied (see main text and Table
  \protect\ref{tbl:xsections}).}


\EPSFIGURE[p] {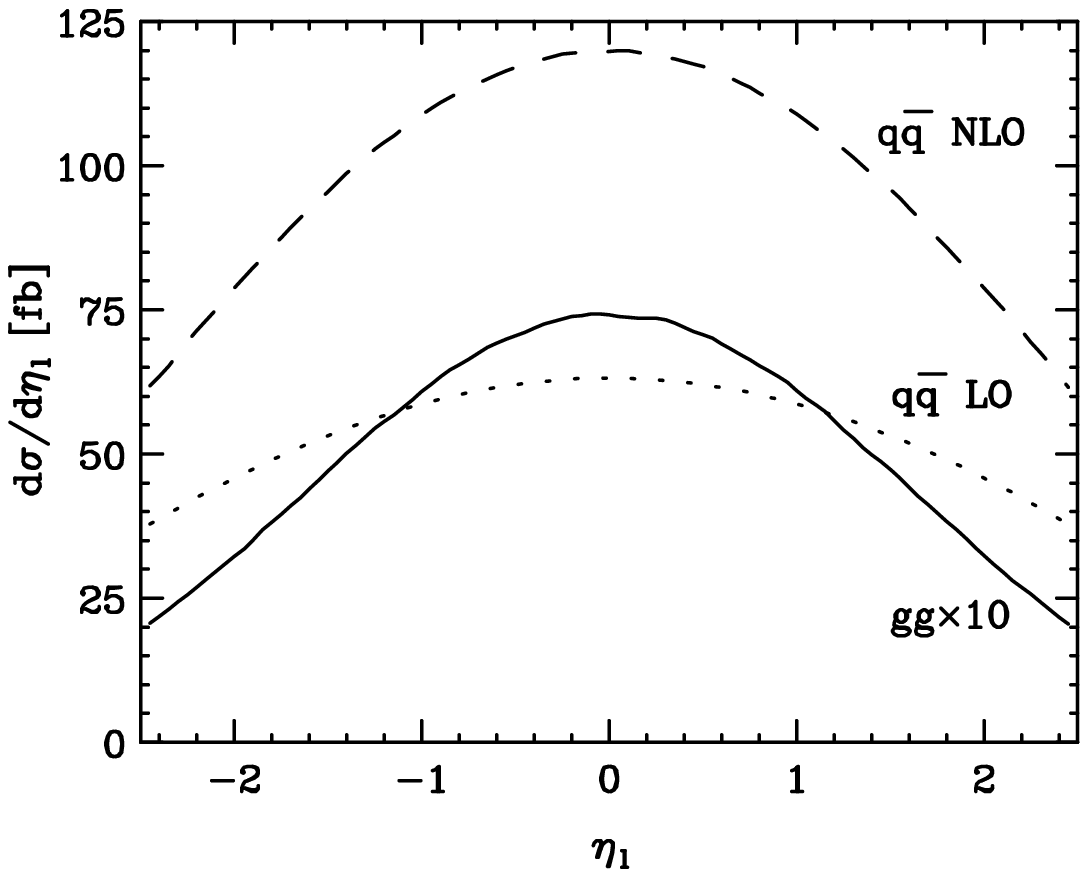,width=10.96cm} { \label{fig:etal}
  Distributions in the pseudorapidity $\eta_{\ell^-}$ of the
  negatively charged lepton. Details as in Fig.~\protect\ref{fig:mll}.
  The $gg$ distribution is displayed after multiplication with a
  factor 10.}


\EPSFIGURE[p] {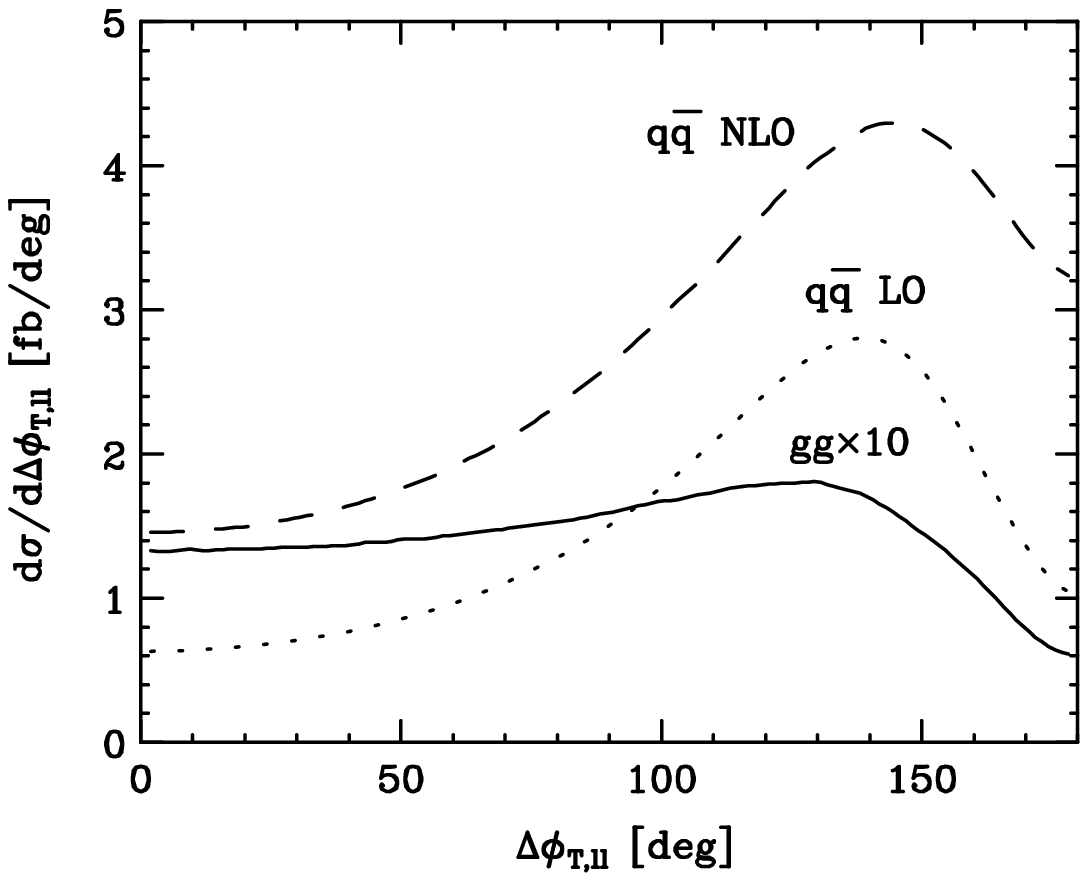,width=11.5cm} {
  \label{fig:delphill} Distributions in the transverse-plane opening
  angle of the charged leptons $\Delta\phi_{T,\ell\ell}$.  Details as in
  Fig.~\protect\ref{fig:mll}. The $gg$ distribution is displayed after
  multiplication with a factor 10.}


\end{document}